\documentclass[useAMS,usenatbib]{mn2e}

\RequirePackage[dvips]{graphicx}

\usepackage{color}
\def\aap{\emph{A.\& A.}}

\def\apj{\emph{ApJ.}}
\def\apjs{\emph{ApJ. Supp.}}
\def\apjl{\emph{ApJ. Lett.}}

\def\mnras{\emph{MNRAS}}

\title[The Minimum Variability Time Scale and its Relation to Pulse Profiles of Fermi GRBs]{The Minimum Variability Time Scale and its Relation to Pulse Profiles of Fermi GRBs}

\author[MacLachlan et al.]{G. A. MacLachlan$^{1}$\thanks{E-mail:
maclach@gwu.edu (GAM)}, A. Shenoy$^{1}$, E. Sonbas$^{2,3}$, K. S. Dhuga$^{1}$, A. Eskandarian$^{1}$, 
\newauthor L. C. Maximon$^{1}$, and W. C. Parke$^{1}$\\
$^{1}$Department of Physics, The George Washington University, Washington, D.C. 20052, USA.\\
$^{2}$University of Adiyaman, Department of Physics, 02040, Adiyaman, Turkey.\\
$^{3}$NASA Goddard Space Flight Center, Greenbelt, MD 20771, USA.\\
}

\begin{document}

\maketitle

\label{firstpage}

\begin{abstract}
We present a direct link between the minimum 
variability time scales extracted through a wavelet decomposition and 
the rise times of the shortest pulses extracted via fits of 34 Fermi GBM GRB light curves
comprised of 379 pulses. Pulses used in this study were fitted with log-normal functions 
whereas the wavelet technique used employs a multiresolution analysis that does not 
rely on identifying distinct pulses.
By applying a corrective filter to published data fitted with pulses we demonstrate agreement
between these two independent techniques and offer a method for distinguishing signal from noise. 

\end{abstract}


\section{Introduction}\label{Introduction}
One approach for probing light curves which has received 
attention~\citep{Nemiroff00,Norris05,Hakkila09,Nemiroff12} 
is to express them as a series of displaced pulses, each with a parametric form.
There is an appeal to this approach because 
fitting routines are well-understood and interpretations of rise time, decay 
time, full width at half max, etc, are possible. On the other hand, one must
make certain assumptions when using the pulse-fitting procedure such as
the choice of the functional form to use for an individual pulse 
and the number of parameters to be included in the fitting function.
Moreover, light with high variability at low
power may show variations which are not statistically significant.

A complementary approach using a wavelet-based analysis of a set of 
both long and short GRB light curves 
is discussed by~\citet{MacLachlan12}
in which a time scale, 
$\tau_\beta$, is identified that marks the transition from white noise to a
power law in the power density spectrum (a $f^{-\alpha}$ behavior).
It is argued that over time scales smaller than
$\tau_\beta$ the light curves appear stochastic and signal power is 
distributed uniformly. At time scales larger than $\tau_\beta$, 
identifiable structures (such as pulses) 
with signal power are no longer distributed uniformly 
over the periods of light variation. For this
reason $\tau_\beta$ is referred to as the \emph{minimum variability
time scale}. 

The analysis presented in~\citep{MacLachlan12} is a non-parametric approach
to probing light curves for time scales. It makes no assumptions about the 
nature of the structures in a given light curve that give rise to the 
$f^{-\alpha}$ character. The technique, however, offers no firm connection 
between $\tau_\beta$ and the constituent structures although it seems 
reasonable to associate  $\tau_\beta$ with the scale of the smallest 
emitting structures present. 
 
Results from an application of a log-normal pulse-fitting procedure 
to GRB light curves
have been published by~\citet{Bhat12}. In this paper we make a meta-analysis 
of the timing results presented by~\citet{Bhat12} 
compared with the techniques 
of~\citet{MacLachlan12} for a set of 34 GRBs used in both 
studies.

\section{Analysis}
We begin by considering the relation between $\tau_\beta$ and the pulse 
parameters given in Table 3 of~\citet{Bhat12}. The parameters with temporal
units in Table 3 are: \emph{time-since-trigger}, \emph{rise time}, 
\emph{decay time}, and \emph{FWHM}. In all, 34 GRBs comprising 379 pulses are
considered here. We note that rise time, decay time, and 
FWHM as presented in  Table 3 of~\citet{Bhat12} are tightly correlated
and for the argument that follows are interchangeable without affecting the 
conclusions. However, we use rise time to make our argument because, as~\citet{Bhat12}
noted, rise times are observed to be shorter than decay time and FWHM (see Table 3 in 
~\citet{Bhat12}).
We considered only those light curves
from NaI detectors and summed over the energy acceptance as in Table 3 of~\citet{Bhat12} 
and in~\citet{MacLachlan12}.  

\begin{figure}
\includegraphics[width=84mm]{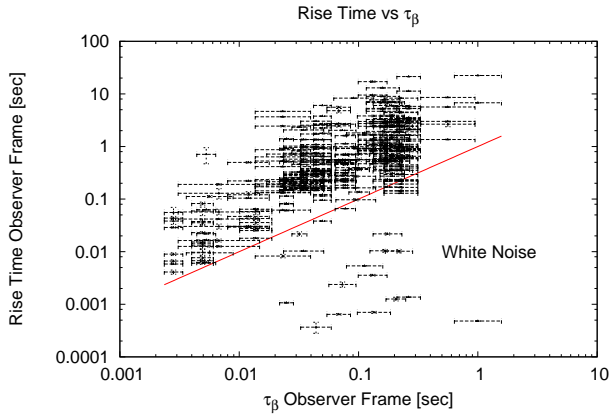}
\caption{Minimum variability time scale versus rise time in the Observer frame.
The rise times are taken from Table 3 of~\citet{Bhat12}. The line represents
the locus where $\tau_\beta$ = rise time. We identify the area below the line 
with white noise. The data are expected to press up against the line from above 
but not to cross it. }
\label{fig:MTS_Rise_ErrsNoCUTS}
\end{figure}

In Fig.~\ref{fig:MTS_Rise_ErrsNoCUTS} we plot the rise time for all 
379 pulses (34 GRBs) along the vertical axis and $\tau_\beta$ along the 
horizontal. Note that for each GRB for which one $\tau_\beta$ is computed,
there is a possibility for multiple pulses and therefore multiple rise times,
hence the vertical columns of rise times for a single value of $\tau_\beta$.
For a given column of pulse times the shortest pulse rise times 
are at the bottom and one finds larger rise times by moving up the column.
An equality line is also shown which is the locus where 
$\tau_\beta$ equals rise time. Arguing as we do that $\tau_\beta$ represents
the minimum variability time scale the space in the $\tau_\beta$-rise time 
plane below the equality line should be interpreted as a structureless 
white noise region. If some method were capable of discerning light curve
structure in the region we define as white noise, then our assertion of 
having found a minimum variability time scale will have been disproven.
Indeed, in Fig.~\ref{fig:MTS_Rise_ErrsNoCUTS} there 
are 27 pulses with rise times below the equality line. The 
uncertainties accompanying these 27 rise times are small, making their
intrusion into the white noise region significant.
\begin{table*}
\centering 
\caption{Pulses with rise times smaller than $\tau_\beta$ but larger than bin widths. Pulse Number (\#), rise (time), $\delta$rise (time), and bin width are taken from Table 3 of~\citet{Bhat12}. The column labeled $\Delta$rise (time) is obtained by combining $\delta$rise and bin width in quadrature. 
The columns diff and \% diff refer to the differences between $\tau_\beta$ and rise (time).}
\begin{tabular}{ccccccccccc} 
\hline 
GRB & Pulse \# & $\tau_{\beta}$ [s] & $\delta\tau^-_{\beta}$ [s]  & $\delta\tau^+_{\beta}$ [s]  &rise [s]& $\delta$rise [s] & $\Delta$rise [s]  & bin width [s]& diff [s]& \% diff \\
\hline
080825593 & 17 & 0.0775 & 0.0138 & 0.0168 & 0.0660 & 0.0003 & 0.0200 & 0.0200 & 0.0115 & 17.4 \\ 
080916009 & 9 & 0.2266 & 0.063 & 0.0872 & 0.1670 & 0.0022 & 0.1500 & 0.1500 & 0.0596 & 35.7 \\ 
080916009 & 15 & 0.2266 & 0.063 & 0.0872 & 0.2190 & 0.0018 & 0.1500 & 0.1500 & 0.0076 & 3.47 \\ 
080916009 & 20 & 0.2266 & 0.063 & 0.0872 & 0.1930 & 0.0016 & 0.1500 & 0.1500 & 0.0336 & 17.4 \\ 
080916009 & 22 & 0.2266 & 0.063 & 0.0872 & 0.2260 & 0.0027 & 0.1500 & 0.1500 & 0.0006 & 0.265 \\ 
080925775 & 10 & 0.1748 & 0.0425 & 0.0562 & 0.1710 & 0.0035 & 0.0501 & 0.0500 & 0.0038 & 2.22 \\ 
081215784 & 1 & 0.0319 & 0.0043 & 0.005 & 0.0218 & 0.0020 & 0.0054 & 0.0050 & 0.0101 & 46.3 \\ 
\hline 
\end{tabular}
\label{table:goodshortrisetimes} 
\end{table*}

\begin{figure}
\includegraphics[width=84mm]{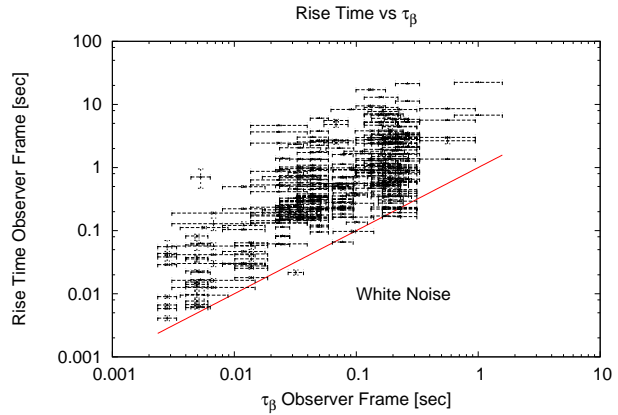}
\caption{Minimum variability time scale versus rise time in the Observer frame.
We have removed all pulses (20) with rise times smaller than the light 
curve bin width. }
\label{fig:MTS_Rise_Errs}
\end{figure}

However, a closer inspection of Table 3 of~\citet{Bhat12} reveals that there
are 20 light pulses in Fig.~\ref{fig:MTS_Rise_ErrsNoCUTS} 
with rise times that are \emph{smaller} than the smallest 
bin widths, in some cases smaller by factors of ten or a hundred. 
Moreover, of those 20 pulses there are 16 pulses 
in Fig.~\ref{fig:MTS_Rise_ErrsNoCUTS} 
with full widths at half max that are \emph{smaller} than the smallest 
bin widths and indeed those 16 all fall below the equality line. 
While it seems that inclusion of these pulses in Table 3 is important for the
sake of completeness, we question the physical reality of these pulses. 
Note that in~\citet{MacLachlan12} all light curves are binned at 
200 microseconds. 
Fig.~\ref{fig:MTS_Rise_Errs} shows the effect of removing the
20 non-physical pulses. 
Note that in Fig.~\ref{fig:MTS_Rise_Errs} the white noise region has been 
vacated by all but seven points and none of the pulse rise times 
above the equality line have been disturbed by the bin width cut. 
For the seven points that remain beneath the equality line, we show in 
Table~\ref{table:goodshortrisetimes} 
that six are within one sigma of the equality line.

\begin{figure}
\includegraphics[width=84mm]{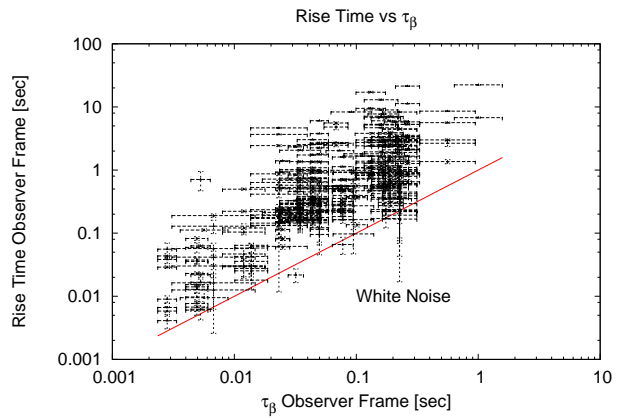}
\caption{Minimum variability time scale versus rise time in the Observer frame
as in Fig.~\ref{fig:MTS_Rise_Errs}. We have folded a single bin width into 
the rise time uncertainties.}
\label{fig:MTS_Rise_CombQuadErrs}
\end{figure}

We make one other point regarding the pulse rise times in 
Fig.~\ref{fig:MTS_Rise_ErrsNoCUTS}
and 
Fig.~\ref{fig:MTS_Rise_Errs}, in particular regarding the 
size of the uncertainties. 
Of the 379 pulse rise times reported by~\citet{Bhat12} and used 
for this meta-analysis, 301 have uncertainties smaller than the 
binning of the light curve, in some cases hundreds or thousands of times 
smaller. We argue that a conservative estimate of the uncertainties for 
the pulse rise times should be no smaller than a bin width. Thus, we add in 
quadrature a bin width (as reported in Table 3 of~\citet{Bhat12}) 
to the rise time uncertainties (also reported in Table 3 
of~\citet{Bhat12}) and plot the result in 
Fig.~\ref{fig:MTS_Rise_CombQuadErrs}.

\begin{figure}
\includegraphics[width=84mm]{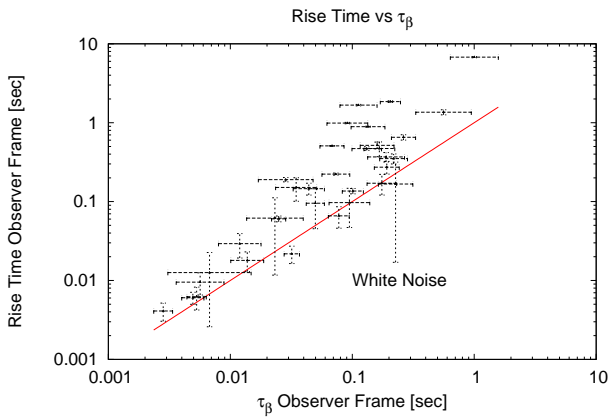}
\caption{Minimum variability time scale versus rise time in the Observer frame
as in Fig.~\ref{fig:MTS_Rise_CombQuadErrs} but with only smallest rise times
included. Note that the equality line between $\tau_\beta$ and rise time 
marks a boundary between scaling processes and white noise and gives substance
to the interpretation of the minimum variability time scale.}
\label{fig:MTS_Rise_CombQuadErrsMin}
\end{figure}

In Fig.~\ref{fig:MTS_Rise_CombQuadErrsMin} we plot only the smallest rise times
for each GRB against $\tau_\beta$. We argue that by rejecting pulse rise times
smaller than light curve bin widths and by folding rise time uncertainties with
a bin width we get strong evidence that \citet{Bhat12} and \citet{MacLachlan12}
have tracked the same physical observables over approximately three orders
of magnitude using independent methods.


\section{Results and Discussion}
For a large sample of short and long Fermi GBM bursts, \citet{MacLachlan12} 
used a technique based on wavelets to determine the minimum variability 
time scale, $\tau_\beta$.
The authors 
associate this time scale with a transition from red-noise processes to parts 
of the power spectrum dominated by white noise or random noise components. 
Accordingly, the authors note that this time scale is the shortest resolvable 
variability time for physical processes intrinsic to the GRB. In addition, histograms of 
the values of $\tau_\beta$ for long and short GRBs were 
shown to exhibit a clear temporal 
offset in the mean $\tau_\beta$ values for long and short GRBs. 

In a separate analysis, using a particular functional form for pulse shapes, 
\citet{Bhat12} have extracted an extensive set of key pulse parameters such 
as rise times, decay times, widths (FWHM), and times since trigger for a host 
of bright GRBs detected by Fermi/GBM. Using the FWHM values, these authors also 
reported a significant temporal offset between the mean values for long and 
short GRBs. 

Although neither group offers a concrete explanation for the temporal 
difference between the distributions of long and short GRBs, it is noteworthy 
that they arrive at a result which is quantitatively in good agreement
with one another, 
especially having used independent approaches. Both sets of analyses also 
suggest scaling trends between characteristic timescales. In the case of 
minimum variability timescales the trend is between $\tau_\beta$ 
and the duration 
of the burst, typically denoted by $T_{90}$. For the pulse-shape analysis, the 
trend is more readily evident and is demonstrated through a number of positive 
correlations involving key parameters such as rise times, decay times and FWHM
times. 

It is relatively straightforward to interpret the scaling trends in terms of 
the internal shock model in which the basic units of emission are assumed to 
be pulses that are produced via the collision of relativistic shells emitted 
by the central engine. In the case of the pulse-fitting method this is 
essentially the default assumption. Indeed, \citet{Quilligan02} in their 
study of the brightest BATSE bursts with $T_{90} > 2$ sec were the first to 
demonstrate this explicitly by identifying and fitting distinct pulses and 
showing a strong positive correlation between the number of pulses and the 
duration of the burst. More recent studies,~\cite{Bhat11,Hakkila08,Hakkila11} 
have provided further evidence for the pulse paradigm view of the prompt 
emission in GRBs. 

The wavelet analysis does not, however, rely on identifying distinct pulses but 
instead uses the multiresolution capacity of the wavelet technique to resolve 
the smallest temporal scale present in the prompt emission. 
Nonetheless, as~\citet{MacLachlan12} have demonstrated, 
if the smallest temporal scale is due to pulse emissions, 
then we can still get a measure of the upper bound on the number of 
pulses in a given burst through the ratio $T_{90}/\tau_\beta$. 
In the simple model in which a pulse is 
produced every time two shells collide, 
the ratio $T_{90}/\tau_\beta$, 
should show a correlation with the duration of the 
burst. Indeed, this correlation was reported by \citet{MacLachlan12}. 

The similar trends of scaling demonstrated by these two methods, not only 
suggest the robustness of both methods, but also point, perhaps more 
importantly, to an underlying interconnection between key parameters extracted 
by the two techniques. In other words, the minimum variability time scale 
extracted by the wavelet 
technique is directly 
related to key pulse time parameters such as rise 
times (as depicted in Fig.~\ref{fig:MTS_Rise_CombQuadErrsMin}), 
under suitably controlled pulse-fitting methods.

\section{CONCLUSIONS}

Through a meta-analysis of results presented by \citet{Bhat12} 
and by \citet{MacLachlan12}, we have studied the relationship between key 
parameters that describe the temporal properties of a sample of 
prompt-emission light curves for long and short-duration GRBs detected by 
the Fermi/GBM mission. We compare the minimum variability timescale extracted 
through a technique based on wavelets, with the pulse-time parameters extracted 
through a pulse-fitting procedure. Our main results are summarized as follows:

a) Both methods indicate a temporal offset between short and long-duration 
bursts. The quantitative agreement between the two methods is quite good.

b) Both methods point to scaling trends between characteristic timescales. 
In the case of the pulse-fitting method the scaling appears to involve 
parameters such as rise times, FWHM, and pulse intervals. For the wavelet 
technique, the scaling involves a correlation between the minimum variability 
time scale and the duration of the bursts.

c) By demonstrating a strong positive correlation between $\tau_\beta$ and the 
rise time of the shortest fitted pulses, we provide for the first time, a 
direct link between the shortest resolvable temporal structure in a GRB 
light curve with that of a key pulse profile parameter. 

d) By combining the two techniques, we have shown that one can arrive at a much tighter 
demarcation of the boundary between the power spectrum domains that separate red noise and white noise processes. 


\section{ACKNOWLEDGEMENTS}

The NASA grant NNX11AE36G provided partial support for this work and 
is gratefully acknowledged. The authors, in particular GAM and KSD, 
acknowledge very useful discussions with Jon Hakkila and Narayan Bhat.

\end{document}